\documentclass[aps,twocolumn,prl,superscriptaddress,preprintnumbers,showpacs,tightenlines]{revtex4}
\usepackage{amssymb}
\usepackage{graphicx,times}

\begin{document}

\preprint{First Draft}
\title{Repeated measurements and nuclear spin polarization}
\author{Lian-Ao Wu}
\affiliation{Department of Theoretical Physics and History of Science, The Basque Country
University (EHU/UPV), PO Box 644, 48080 Bilbao, Spain}
\affiliation{IKERBASQUE, Basque Foundation for Science, 48011 Bilbao, Spain}
\pacs{03.67.-a, 73.21.La, 76.70.-r}

\begin{abstract}
We study repeated (noncontinuous) measurements on the electron spin in a
quantum dot and find that the measurement technique may lead to a different method 
or mechanism to realize nuclear spin polarization.
While it may be used in any case, the method is aimed at the
further polarization, providing that nuclear spins have been polarized by
the existent electrical or optical methods. The feasibility of the method is
analyzed. The existing techniques in electron spin measurements are
applicable to this scheme. The repeated measurements \emph{deform} the
structures of the nuclear wave function and can also serve as $\emph{gates}$
to manipulate nuclear spins.
\end{abstract}

\maketitle

\textit{Introduction.--- }Quantum dots can host electron spin qubits in a
quantum information processor \cite{Loss98,Burkard00,Kane98,Wu04}, as
evidenced by the recent encouraging progress in spin detection, relaxation
and coherent manipulation \cite{Elzerman04,Sherman05,Petta05}. However,
random fluctuations in the nuclear spin ensemble of a host quantum dot
lead to fast electron spin decoherence \cite{Khaetskii02,Merkulov02} via the
hyperfine coupling. Methods to combat the nuclear spin randomness, nuclear
spin polarization, have been proposed (see \cite{Imamoglu03,Wu10} and
references therein) and implemented experimentally.

Nuclear spin polarization dates back to 1980s \cite{Meier84}. It has been
realized experimentally to some extent. The polarization is achieved either
electrically or optically via the hyperfine coupling in particular flip-flop
spin exchange \cite{Imamoglu03}. Gammon \textit{et al.} obtain 60 \% nuclear
spin polarization optically in interface fluctuation GaAs quantum dots \cite%
{Gammon96} (see also \cite{Hu10} and references therein). The recent record
is the 80\% polarization in In$_{0.9}$Ga$_{0.1}$As at 5T \cite{Maletinsky08}.

Although significant progress has been made, the full nuclear spin
polarization remains far from reach. Nuclear spins distant from the electron
spin are subject to rather weak hyperfine coupling \cite{Yao09}. Further
electrical and optical polarization becomes more difficult, even impossible.
However, proposals to use quantum dots for quantum information processing
are based on the fully polarized state \cite{Taylor03,Wu10}. Partial
polarizations do not yet lead to a significant application in quantum
information processing. Therefore, new polarization methods are desired.

Existent polarization methods, both electrical and optical, are dynamic.
Here, we explore an alternative possibility of nuclear spin polarization via
frequent but $\emph{noncontinuous}$ measurements. We start with an achieved
80\% \emph{polarized} equilibrium state. If we measure the electron spin
repeatedly but noncontinuously ( different from the Zeno effect \cite{Loss08}
) as formulated in \cite{Nakazato03}, the system may end up with the fully
polarized nuclear spin state. While the hyperfine coupling is required in
this scheme, the relative stronger nuclear spin correlation plays a positive
role. This feature may favor the polarization of those remote nuclear spins
subject to weaker hyperfine coupling.

\textit{Method.---} We consider a single electron confined in a charged
quantum dot. The Hamiltonian for the electron spin and its surrounding $K=10^{3}-10^{5}$ 
nuclei spins ( spin-$I$ ) is 
\begin{equation}
H=g^{\ast }\mu _{B}BS_{z}+g_{n}\mu _{n}BI_{z}+H_{I}+H_{nuc},  \label{total}
\end{equation}%
where the first two terms are the Zeeman energies of the electron spin and
nuclear spins in a \textit{z} direction magnetic field $B.$ $%
I_{z}=\sum_{i}I_{z}^{i}$ is the \textit{z} component of the total nuclear
spin operator. The total angular momentum operator in the $z$ direction, $%
J_{z}=S_{z}+$ $I_{z}$ with eigenvalue value $J$, is conserved. The hyperfine
coupling between nuclear spins and the electron spin is written as%
\begin{equation}
H_{I}=\mathcal{A}(A_{z}S_{z}+\frac{1}{2}A_{+}S_{-}+\frac{1}{2}A_{-}S_{+}),
\label{coupling}
\end{equation}%
where $\mathcal{A}/\sqrt{2I}$ is an average hyperfine coupling constant.
Operators $A_{\mu }=\sum_{i}\alpha _{i}I_{\mu }^{i}/\sqrt{2I}$ are expressed
by the nuclear spin $I_{\mu }^{i}$ ( $\mu =z,+,-$ ), where the real numbers $%
\alpha _{i}$'s satisfy $\sum_{i=1}^{K}\alpha _{i}^{2}=1$ \cite{Wu10}. The
term $A_{z}S_{z}$ provides an effective magnetic field \cite{Taylor03},
Overhauser shift for the electron spin, $B_{eff}=B-(g_{n}\mu _{n}B+\mathcal{A%
}\sum \alpha _{i}I_{z}^{i})/g^{\ast }\mu _{B}$. The flip-flop spin exchange $%
\frac{1}{2}A_{+}S_{-}+\frac{1}{2}A_{-}S_{+}$ has been used in either
electrical or optical nuclear spin polarizations \cite{Imamoglu03,Hu10}. The
dominate part of nuclear spin interaction $H_{nuc}$ is the dipole-dipole
coupling. However, here we consider the most general nuclear spin
interaction $H_{nuc}$ \cite{Hu10} preserving the total nuclear spin $I_{z}$%
. This general Hamiltonian can be represented by a direct sum of submatrices 
$\mathbf{H}_{J}$, in the ordered bases $\left\{ \left\vert J\right\rangle
\right\} $ (from $J=-J_{m}$ to $J_{m}$) 
\[
\mathbf{H}=\sum_{J=-J_{m}}^{J_{m}}\oplus \mathbf{H}_{J},
\]%
where $J_{m}=KI+1/2$. Given numbers $J$ and $I$, there are $\Omega (I,N)$
states in the electron and nuclear spin Hilbert space, where $N=(J_{m}-J)/I$%
. For instance, $\Omega (1/2,N)=\frac{(K+1)!}{(K+1-N)!N!}$ when $I=1/2$. The
representation $\mathbf{H}_{J_{m}}$ is one-dimensional, whose basis $%
\left\vert J_{m}\right\rangle =\left\vert \Uparrow \right\rangle \left\vert 
\mathbf{0}\right\rangle $ ( $\left\vert \mathbf{0}\right\rangle =\left\vert
I,I,...,I\right\rangle $ ) is the fully polarized electron and nuclear spin
state. Here $\left\vert \Uparrow \right\rangle $ denotes the electron
spin-up state and $\left\vert \mathbf{0}\right\rangle =\left\vert
I,I,...,I\right\rangle $ is the fully polarized state of nuclear spins.

Consider the electron spin and its surrounding nuclear spins initially in a
separable state $\rho (0)=\left\vert \Uparrow \right\rangle \left\langle
\Uparrow \right\vert \otimes \rho _{n}$, where $\rho _{n}$ is an initial
nuclear spin state. This initial state can be obtained by measuring the
electron at $t=0$ with output $\left\vert \Uparrow \right\rangle $ - the
spin-up state. The system evolves under the Hamiltonian (\ref{total})
thereafter. We then measure the electron spin again at time $t=\tau $. If we
find\ again that the electron is in spin-up state, which we call a \emph{%
successful} measurement, the nuclear spin state becomes $\rho _{n}(\tau
)=V(\tau )\rho _{n}V^{\dagger }(\tau )/P_{1}$, where $V(\tau )=\left\langle
\Uparrow \right\vert \exp (-iH\tau )\left\vert \Uparrow \right\rangle $.
After\ $M$ such \emph{noncontinuous} measurements on electron spin state,
given that all outcomes were $\left\vert \Uparrow \right\rangle $, the
nuclear spin state \cite{Wu041} becomes%
\begin{equation}
\rho _{n}(M\tau )=V(\tau )^{M}\rho _{n}V^{\dagger }(\tau )^{M}/P_{M},
\label{evolution}
\end{equation}%
where $P_{M}=$Tr$[V(\tau )^{M}\rho _{n}V^{\dagger }(\tau )^{M}]$. The matrix
representation of the operator $V(\tau )$ is 
\begin{equation}
\mathbf{V}(\tau )=\left[ 
\begin{array}{cc}
e^{-iE_{J_{m}}\tau } & \mathbf{0} \\ 
\mathbf{0} & \mathbf{V}_{r}(\tau )%
\end{array}%
\right]   \label{V(tau)}
\end{equation}%
where 
\[
\mathbf{V}_{r}(\tau )=\sum_{I_{z}=-KI+1}^{KI}\oplus \mathbf{V}_{I_{z}}(\tau )
\]%
in the ordered basis $\left\{ \left\vert I_{z}\right\rangle \right\} $. $%
V(\tau )$ behaves like a non-Unitary gate in the evolution. These
measurements preserve the total nuclear spins, $I_{z}=J-1/2$. We emphasize
that measurements are noncontinuous, for the interval $\tau $ is finite and
could be long (different from the Zeno effect in quantum dot \cite{Loss08}).
The submatrix of $\mathbf{V}(\tau )$ for the fully polarized nuclear state
is also one-dimensional, where $E_{J_{m}}$ is the eigenenergy of $H$.
Matrices $\mathbf{V}_{I_{z}}(\tau )$'s are in general neither Hermitian nor
Unitary. However, one can still find left- and right-eigenvectors with
complex eigenvalues, whose modulus can be shown to be bounded between $0$
and $1$ \cite{Wu041}. When $M$ goes to infinite, $\rho _{n}(M\tau
)\rightarrow \left\vert \mathbf{0}\right\rangle \left\langle \mathbf{0}%
\right\vert $, the nuclear spins are fully polarized, if and only if the
moduli of all eigenvalues $v_{I_{z}}(\tau )$ of $\mathbf{V}_{I_{z}}(\tau )$
are smaller than one, i. e., $\left\vert v_{I_{z}}(\tau )\right\vert <1$ for
all $I_{z}<KI.$ However, if there are other eigenstates with moduli equal to
one in the submatrix $\mathbf{V}_{I_{z}}(\tau )$, $\rho _{n}(M\tau )$ will
end up with a \emph{degenerate} mixed state of the fully polarized state and
these eigenstates. The feasibility of this polarization method relies on
whether or not there is \emph{degeneracy} of $V(\tau )$ for the general
Hamiltonian (\ref{total}). Since a complete solution for the general Hamiltonian is
impossible, the following sections will discuss several limits, numerically
and analytically, to verify that there is no \emph{priori} reason for the
general Hamiltonian to have the \emph{degeneracy}. Before proceeding it, we
first introduce the initial nuclear spin state $\rho _{n}$, whose nature
also plays a role in this polarization method.

\textit{Initial state.---} Consider that the nuclear spins have been
polarized initially to a polarization rate $a$ ( e.g. equal to $80\%$ ),
either electrically or optically. An initial state can be a product of
individual spin states $\rho _{n}=\prod_{i=1}^{K}\left[ a\left\vert
I\right\rangle _{i}\left\langle I\right\vert +(1-a)\eta _{i}\right] $, \
meaning that each nuclear spin is in its \emph{evenly} \emph{polarized}
equilibrium state. Here $\eta _{i}$ is a normalized state orthogonal to $%
\left\vert I\right\rangle _{i}\left\langle I\right\vert $. For simplicity,
we set $K$ even and $I=1/2$ such that $\eta _{i}=\left\vert
-1/2\right\rangle _{i}\left\langle -1/2\right\vert .$ We expand this initial
state as a direct sum,%
\[
\rho _{n}=\sum_{I_{z}=-KI}^{KI}\oplus c(I_{z},a)\rho _{n}^{I_{z}}
\]%
where $c(I_{z},a)=d_{I_{z}}a^{K/2+I_{z}}(1-a)^{K/2-I_{z}}$ and $d_{I_{z}}=%
\frac{K!}{(K/2-I_{z})!(K/2+I_{z})!}$. Here the density matrix $\rho
_{n}^{I_{z}}=(1/d_{I_{z}})$diag$(1,1,...,1)$ is a $d_{I_{z}}$- dimensional
diagonal and normalized matrix in the basis $\left\{ \left\vert
I_{z}\right\rangle \right\} $. In the maximally mixed state $(a=1/2$ or $50\%
$ $)$, the density matrix $\rho _{n}^{I_{z}}$ with $I_{z}=0$ is dominant.
However, it will be different for a polarized state. The ratio between
coefficients of $I_{z}=0$ states and fully polarized state $\left\vert 
\mathbf{0}\right\rangle \left\langle \mathbf{0}\right\vert $ is $R=\frac{%
K!(1-1/a)^{K/2}}{(K/2)!^{2}},$ for instance $K=10^{5}$ , $R$ is $2.5\times
10^{-3}$ when $a=80\%$. The fully polarized state $\left\vert \mathbf{0}%
\right\rangle \left\langle \mathbf{0}\right\vert $ is dominant. The
expectation value of $I_{z}$ characterizing the polarization degree is%
\begin{equation}
\left\langle I_{z}\right\rangle _{M}=\sum_{I_{z}=-KI}^{KI}\frac{%
c(I_{z},a)I_{z}}{d_{I_{z}}P_{M}}\text{Tr}[\mathbf{V}_{I_{z}}(\tau )^{M}%
\mathbf{V}_{I_{z}}^{\dagger }(\tau )^{M}],  \label{expect}
\end{equation}%
where the trace runs over all subspaces characterized by $I_{z}.$ Initially,
the expectation value $\left\langle I_{z}\right\rangle _{0}=K(a-1/2)$ and a
nuclear spin polarization process means that $\left\langle
I_{z}\right\rangle _{M}\rightarrow K/2$ when $M$ increases.

There is another possible initial state, where the eighty percent of nuclear
spins is polarized. The other $20\%$ of spins is distant from the electron
and is in thermal equilibrium state ($a=1/2$). It is an \emph{uneven} \emph{%
polarized} state. The expectation value of $I_{z}$ for this initial state is 
\begin{equation}
\left\langle I_{z}\right\rangle _{M}=aKI+\left\langle I_{z}\right\rangle
_{M}^{\prime },  \label{expect1}
\end{equation}%
where $\left\langle I_{z}\right\rangle _{M}^{\prime }$ has the same form as (%
\ref{expect}) but $I_{z}$ in the sum run from $-(1-a)KI$ to $(1-a)KI$.

\bigskip \textit{The flip-flop spin exchange.---} Eigenstates $\left\vert
m\right\rangle $ of a Hermitian operator $h_{0}=A_{+}A_{-}$ play crucial
roles for the dressed qubit supported by $\left\vert 0\right\rangle
_{d}=\left\vert \Downarrow \right\rangle \left\vert m\right\rangle $ and\ \ $%
\left\vert 1\right\rangle _{d}=\left\vert \Uparrow \right\rangle \left\vert
\Phi _{m+1}\right\rangle $ \cite{Wu10}. These two states span an invariant
space of the flip-flop spin exchange.\ Note that here we define the state $%
\left\vert \mathbf{0}\right\rangle =\left\vert I,I,...,I\right\rangle $ as
our \emph{vacuum} state instead of $\left\vert -I,-I,...,-I\right\rangle $
in ref. \cite{Wu10}. Ref. \cite{Wu10} also uses $K$ operators, $%
A_{k-}=\sum_{i}\alpha _{i}^{k}I_{-}^{i}/\sqrt{2I}$, and identifies the
collective mode $k=0$ such that $A_{-}=A_{0-}$ and $\alpha _{i}=\alpha
_{i}^{0}$. The set $\{\alpha _{i}^{k}\}$ can be made as a unitary matrix 
\cite{Kurucz09,Wu10} to transform sites $i$ to modes $k$. These operators
obey the commutation relations%
\begin{equation}
\lbrack A_{k+,}A_{k^{\prime }-}]=\delta _{kk^{\prime }}-\sum_{i}\alpha
_{i}^{k\ast }\alpha _{i}^{k^{\prime }}(I-I_{z}^{i})/I.  \label{commutor}
\end{equation}%
An eigenstate $\left\vert m\right\rangle $ of $h_{0}$ can be expressed by a
polynomial of product states of the operators $A_{k-}$'s, $\left\vert
m\right\rangle =f_{m}(A_{-},A_{1-},...,A_{K-1-})\left\vert \mathbf{0}%
\right\rangle .$ When $f$ do not include the collective operator $A_{-},$
the states $\left\vert m_{0}\right\rangle
=f_{m_{0}}(A_{1-},...,A_{K-1-})\left\vert \mathbf{0}\right\rangle $ are
always eigenstates of $h_{0}$, and $A_{+}\left\vert m\right\rangle _{0}=0$.
For instance, in the $I_{z}=KI-1$ subspace, the corresponding eigenstates
are $\left\vert m\right\rangle =\left\vert \mathbf{0}\right\rangle $ and $%
\left\vert m_{0}\right\rangle =\left\vert \mathbf{1}_{k}\right\rangle
=A_{k-}\left\vert \mathbf{0}\right\rangle $.

Now we come to illustrate natures of the operator $V(\tau ).$ When $%
B_{eff}=0 $ and the nuclear interaction $H_{nuc}$ is negligible, by
calculating $\left\langle \Uparrow \right\vert H^{n}\left\vert \Uparrow
\right\rangle $ and $\left\langle \Uparrow \right\vert e^{-iH\tau
}\left\vert \Uparrow \right\rangle $ we show that 
\begin{equation}
V(\tau )=\cos (\frac{\mathcal{A}\tau \sqrt{h}}{4}),  \label{v1}
\end{equation}%
where $h=A_{-}A_{+}$. While it is easy to check that $h\left\vert \mathbf{0}%
\right\rangle =0,$ the abovementioned states $\left\vert m\right\rangle _{0}$
are also eigenstates with zero eigenvalues, i. e., $h\left\vert
m\right\rangle _{0}=0$. In the $I_{z}=KI-1$ subspace, $h\left\vert \mathbf{1}%
_{k}\right\rangle =0.$ After $M$ measurements with output $\Uparrow $, both $%
\left\vert \mathbf{0}\right\rangle $ and the states $\left\vert
m\right\rangle _{0}$ are project out. This \emph{degeneracy }makes the full
polarization invalid, though the outcoming nuclear state may still be
interesting.

If we consider a finite effective magnetic field $B_{eff}$, a tedious
calculation shows that 
\begin{equation}
V(\tau )=\cos (\frac{\mathcal{A}\tau \sqrt{h}}{4})e^{-i\frac{g^{\ast }\mu
_{B}B_{eff}\tau }{2}}.  \label{v2}
\end{equation}%
The magnetic field does not break the degeneracy either. Therefore, in our
method the flip-flop spin exchange cannot fully polarize the nuclear spins,
though it is crucial in the electrical and optical polarization methods. The
reason is that the flip-flop spin exchange only contains the collective mode 
$k=0$.

\textit{General analysis.---} The general Hamiltonian (\ref{total}) includes
all modes $k=1,...,K-1$. We can also write it in a way different from eq. (%
\ref{total}) 
\begin{equation}
H=\frac{\mathcal{A}}{2}(A_{+}S_{-}+A_{-}S_{+})+\mathcal{H},  \label{general1}
\end{equation}%
where $\mathcal{H}=(g_{n}\mu _{n}B-g^{\ast }\mu _{B}B)I_{z}+\mathcal{A}%
A_{z}(J-I_{z})+H_{nuc}$ acts only on nuclear spins. $J_{z}$ is conserved and
has been replaced by a constant $J$. There does not exist an analytical form of 
$V(\tau )$ for a general $\mathcal{H}$. However, we may find out its natures 
in terms of approximation methods, such as perturbation expansions. If
the measurement time interval $\tau $ is short, we can expand $\left\langle
\Uparrow \right\vert e^{-iH\tau }\left\vert \Uparrow \right\rangle $ in the
power of $\tau $. It is tedious but straightforward to calculate, 
\begin{equation}
\left\langle \Uparrow \right\vert H^{2m}\left\vert \Uparrow \right\rangle
=\sum_{k=0}^{m}T(\mathcal{H}^{2k}),  \label{expand}
\end{equation}%
where $T(\mathcal{H}^{2k})=\sum \mathcal{P}(\mathcal{H}%
^{2k}(A_{-}A_{+})^{m-k})$ and the operation $\mathcal{P}$ is a permutation,
for instance, $\mathcal{P}(\mathcal{H}^{2}A_{-}A_{+})=\mathcal{A}^{2}%
\mathcal{H}A_{-}A_{+}\mathcal{H}$. The sum runs over all possible
permutations. Similarly, we also have an expansion for $\left\langle
\Uparrow \right\vert H^{2m+1}\left\vert \Uparrow \right\rangle $, e. g., $%
\left\langle \Uparrow \right\vert H^{3}\left\vert \Uparrow \right\rangle =%
\mathcal{HA}^{2}A_{-}A_{+}+\mathcal{A}^{2}A_{-}A_{+}\mathcal{H}+\mathcal{A}%
^{2}A_{-}\mathcal{H}A_{+}+\mathcal{H}^{3}$. The nuclear Hamiltonian $%
\mathcal{H}$ breaks the degeneracy of $V(\tau )$. For the states $\left\vert
m\right\rangle _{0}=\left\vert \mathbf{1}_{k}\right\rangle $, we can see
directly that the fourth order term $\mathcal{H}A_{-}A_{+}\mathcal{H}\ $\
breaks the degeneracy. $\mathcal{H}$ rotates $A_{k-}$ to a superposition of
other operator $A_{k^{\prime }-}$, including the collective operator $A_{-}$
such that $\mathcal{H}A_{-}A_{+}\mathcal{H}$ has an finite expectation
value. By using the common eigenstates $\left\vert m\right\rangle $ of $%
h_{0} $ and $I_{z}$, we expect that $\left\langle m\right\vert \mathcal{H}%
A_{-}A_{+}\mathcal{H}\left\vert m\right\rangle $ is finite for a general $%
\mathcal{H}$.

\textit{Analysis via Bosonization.---} Consider bosons $B_{k}^{\dagger
}=\sum_{i}\alpha _{i}^{k}b_{i}^{\dagger }$, where the set $\left\{ \alpha
_{i}^{k}\right\} $ is the same as that for nuclear spins. We denote $B=B_{0}$ 
for the collective bosonic mode. $B_{k}$ and $B_{k}^{\dagger }$ obey 
the bosonic commutation relations
\begin{equation}
\lbrack B_{k},B_{k^{\prime }}^{\dagger }]=\sum_{i}\alpha _{i}^{k\ast }\alpha
_{i}^{k^{\prime }}=\delta _{kk^{\prime }}.  \label{commutor1}
\end{equation}%
By comparison with Eq. (\ref{commutor}), the nuclear spin ensemble behaves
like that of collective bosons, $A_{k+}\Leftrightarrow B_{k}$, $A_{k^{\prime
}-}\Leftrightarrow B_{k}^{\dagger }$ and $\left\vert \mathbf{0}\right\rangle
\Leftrightarrow \left\vert 0\right\rangle _{B}$, when nuclear spins satisfy
the condition $\sum_{i}\alpha _{i}^{k\ast }\alpha _{i}^{k^{\prime
}}(I-I_{z}^{i})\ll I$. The total nuclear spin $I_{z}\Leftrightarrow (K-N)I$,
where $N$ is the total boson number. The bosonization of nuclear spin
Hamiltonian is $\mathcal{H}=\sum_{k,k^{\prime }=0}^{K-1}g_{kk^{\prime
}}B_{k}^{\dagger }B_{k^{\prime }},$ where $g_{kk^{\prime }}$ are linear
combinations of parameters in (\ref{total}), under the Holstein-Primakoff
approximation \cite{Kurucz09}. The bosonized $h_{0}=BB^{\dagger }$ and $%
h=B^{\dagger }B$ commute. The common eigenstates of $h_{0}$ and $h$ are the
bosonic Fock states $\left\vert \mathbf{n}\right\rangle =\left\vert
n_{0},n_{1},..,n_{K-1}\right\rangle $, with the vacuum state $\left\vert 
\mathbf{n}=\mathbf{0}\right\rangle =\left\vert \mathbf{0}\right\rangle .$
When $n_{0}=0,$ the Fock states $\left\vert 0,n_{1},..,n_{K-1}\right\rangle $
correspond to $\left\vert m\right\rangle _{0},$ which are eigenstates of the
bosonized $V(\tau )=\cos (\frac{\mathcal{A}\tau \sqrt{B^{\dagger }B}}{4})$,
with eigenvalues 1. Other eigenvalues of $V(\tau )$ are $v_{n_{0}}(\tau
)=\cos (\frac{\mathcal{A}\tau \sqrt{n_{0}}}{4}).$ As long as properly
control time $\tau $ such that $v_{n_{0}}(\tau )<1\,$, we can eliminate the corresponding
states $\left\vert n_{0}\neq 0,n_{1},..,n_{K-1}\right\rangle $
after $M$ successful measurements. The bosonization gives us a more
transparent physical picture.

Assume that $\mathcal{H}$ is small by comparing with the spin flip-flop
interaction, the dominant part of eigenstates of $V(\tau )$ is the Fock states 
$\left\vert \mathbf{n}\right\rangle$. In the Fock state basis, $V(\tau )$ may be diagonal when we
consider the first order perturbation, i. e. 
\begin{equation}
V(\tau )=\sum V_{\mathbf{n}}(\tau )\left\vert \mathbf{n}\right\rangle
\left\langle \mathbf{n}\right\vert .  \label{dia}
\end{equation}%
The above discussion shows that there is no priori reason for the general
Hamiltonian to have $\left\vert V_{\mathbf{n}}(\tau )\right\vert =1$ for all 
$\mathbf{n}\neq \mathbf{0}$.

\textit{Numerical simulations.---} Assume that $V(\tau )\ $can be written in
a diagonal form as in (\ref{dia}), in a certain basis. We can have a rough
estimation by taking $V_{\mathbf{n}}(\tau )$ as an average constant $%
\overline{V_{\mathbf{n}}(\tau )}<1$. With this assumption, we calculate $%
\left\langle I_{z}\right\rangle _{M}$ for quantum dots with $K=10^{3}$. When 
$V_{\mathbf{0}}(\tau )=1$ and $\overline{V_{\mathbf{n}}(\tau )}=0.9$, we
reach the full polarization ($\left\langle I_{z}\right\rangle _{M}=K/2$)
with $M=1100$ successful measurements, initially from the even polarized
state. The reason of requiring such large $M$ is that the component of the fully polarized state
is small in the initial state. The probability $P_{M}$ to find $\Uparrow $ decays fast with $M$ and
becomes extremely low when $M=1100$. For instance, $P_{M}$ is already $0.1$
when $M=11.$ Suppose that we find $\Uparrow $ at $M=11$ and stop the measurement. The
nuclear spins end up with a state $\rho _{n}^{(1)}$, which is more polarized
than $\rho _{n}$. We then start the
repeated measurements again but with the normalized $\rho _{n}^{(1)}$. The
probability is $0.1$ again when $M=11$ for the new round. The very long
coherence time of nuclear spins \cite{Taylor03} has implied that nuclear
spin correlation may not disturb the process. We can repeat the same steps again 
and again until the nuclear spins are fully polarized. It is important that each 
successful step improves the polarization
degree of nuclear spins. We can also prepare thousands of copies and repeatedly measure
them individually. 

Starting from the \emph{uneven} polarized state, we also calculate 
$\left\langle I_{z}\right\rangle _{M}$ in (\ref{expect1}) and find that when 
$M=700$, it is already fully polarized. The value of $M$ depends on the
assumption of initial states significantly. The discussions for the even
polarized initial state are also applicable for the uneven state.

It is not practical to attempt to treat the general Hamiltonian exactly
(even numerically ) since the configurations for a given total $J_{z}$ are
huge. Figure 1 shows a simple numerical case. It is clear that
the nuclear spin polarization increases monotonically with $M$. 
\begin{figure}[th]
\includegraphics[width=8.5 cm]{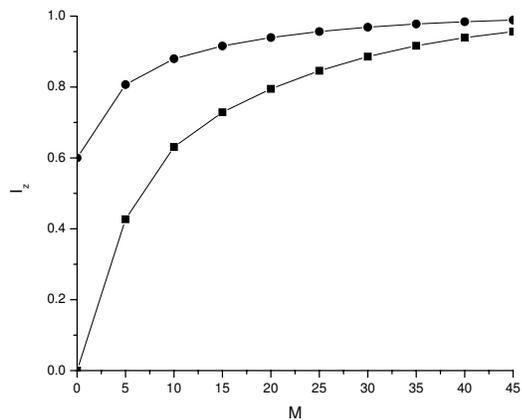}
\caption{A model quantum dot with two nuclear spins. The total nuclear spin 
$\left\langle I_{z}\right\rangle $ vs. $M$ successful measurements. The line
with dots is for the thermal equilibrium initial state $a=1/2$ and the line
with squares is for the 80\% polarized state, $a=4/5.$ In both cases, we
take parameters $\mathcal{A}\protect\alpha _{1}\protect\tau =8,\mathcal{A}
\protect\alpha _{2}\protect\tau =4$ and $b_{12}\protect\tau =0.2$ in the
general Hamiltonian, where $b_{12}$ is the coupling constant of nuclear 
dipole-dipole interaction. Nuclear spins are fully polarized for both 
cases after 50 successful measurements. }
\end{figure}

\bigskip \textit{Measurements.---} Measurement of single electron spin has
been achieved using optical \cite{Blatt88} and electrical \cite{Elzerman04}
techniques. The electrical one \cite{Elzerman04} applies a magnetic field to
split the spin-up and spin-down states of the electron via the Zeeman
energy. The quantum dot potential is then tuned. If its spin is up, the
electron will stay in the dot, otherwise it will leave. The spin state is
correlated with the charge state, and the charge on the dot is measured to
indicate the original spin. This technique matches the present polarization
method. We could exactly follow and repeat the stages of this technique. We
first empty the dot and then inject one electron with unknown spin. At time $%
t=0$, we measure the spin states. If find spin-up, we start processing.
After waiting for time $\tau $, we measure again and if the electron is
spin-up, we continue processing. If find spin-down, we inject an electron
again and go back to time $t=0$. By repeating this, we will end up with the
fully polarized state if we have $M$ successful measurements. Our analysis has
shown that we needs several hundred successful measurements to accomplish
the full polarization. However, it is encouraging that each measurement
with output $\Uparrow$ will increase polarization degrees, as illustrated
in Fig. 1 and discussed above. The full polarization may be reached by
several uncorrelated measurement procedures in the long coherence period of
nuclear spins \cite{Taylor03}.

\textit{Conclusion.--- }Our method is aimed at further polarization, on the
basis of the partially polarized states made by the electrical or optical
methods. Theoretically, the feasibility of the method relies on eigenvalues of $V(\tau )$.
While the modulus of the eigenvalue of the fully polarized state is always
one, we require the\ moduli of other eigenvalues are smaller than one, i.
e., breaking degeneracy of $V(\tau )$. We analyze the system in several
ways. It is interesting to note that while the interaction between nuclei can limit the
polarization process in other polarization schemes \cite{Cirac}, it plays a
positive role on breaking the degeneracy of $V(\tau )$. Our conclusion is
that there is no priori reason that there are other eigenstates with
eigenvalues one for the general Hamiltonian. It is noticeable that the
existent spin measurement technique fits well with our method. This work also
hints that while we explore the feasibility of the polarization
method, our study should surely promote
the development in applying the noncontinuous measurement scheme 
to the interesting nuclear spin system, as a new tool.

The author thanks Drs. E. Sherman, X. Hu and W. Yao for helpful discussions.
This work was supported by the Ikerbasque Foundation and the Spanish MEC
(Project No. FIS2009-12773-C02-02).

\end{document}